\documentclass[12pt,showpacs,showkeys]{revtex4}

\usepackage{amssymb}
\usepackage{amsmath}
\usepackage{graphicx}
\begin{document}
\title{Collective dynamics in coupled maps on a lattice with quenched disorder}

\author{Achille Giacometti $^{(1)}$}
\author{Maurice Rossi $^{(2)}$}
\author{Libero Battiston $^{(3)}$}

\affiliation{$^{(1)}$
INFM Unit\'a di Venezia, Dipartimento
di Chimica Fisica, Universit\'a di Venezia,\\
Calle Larga Santa Marta DD 2137, I-30123 Venezia-Italy}
\affiliation{$^{(2)}$
Laboratoire de Mod\'elisation en M\'ecanique
CNRS Universit\'e Pierre et Marie Curie \\
4, Place Jussieu, 75005 Paris
}
\affiliation{$^{(3)}$
ISAC CNR , Bologna, Unit\'a staccata di Padova
Corso Stati uniti 4, 35100 Padova}
\date{\today}

\begin{abstract}
It is investigated how a spatial quenched disorder
modifies  the dynamics of coupled map lattices.
The disorder is introduced via the presence or absence of coupling
terms among lattice sites. Two  nonlinear maps
have been considered embodying two paradigmatic  dynamics. The Miller and
Huse map can be associated with an Ising-like dynamics, whereas the
logistic coupled maps is a prototype of a non trivial collective
dynamics. Various indicators quantifying the overall behavior, demonstrates that
even a small amount of spatial disorder is capable to  alter the dynamics found
for  purely ordered cases.
\end{abstract}

\maketitle
\newpage

\section{Introduction}
\noindent
Far-from-equilibrium systems, in which the simultaneous presence of non-linear terms and spatial extent is an essential  feature,
involve many important physical or chemical phenomena as varied as Rayleigh-B\'enard
convection in large  horizontal boxes \cite{Morris93},
chemical reaction-diffusion systems \cite{Ouyang96}, colonies of micro-organisms \cite{Lee96}
and fibrillating heart tissue \cite{Qu97}.
In all these examples, the spatio-temporal chaos arising from the coupling of a large number of degree of freedoms, poses
challenging  theoretical questions which has so far eluded a universally accepted
description. Progresses have often been limited  to  models containing the essential ingredients of typical extended  systems
at the crudest possible level. In this respect,  "simple" partial differential equations
such as,  the Kuramoto-Shivashinsky,   Ginzburg-Landau   and   Swift-Hohenberg equations,  have proven to be useful models
to identify   main characteristics  such as symmetry breaking, phase transition, phase dynamics, spatial chaos or defects.
From a numerical point of view, however, these approaches, albeit successful, are rather demanding
in terms of computational time.
Coupled map lattices (CML) \cite{Kaneko93}, where time and space are discretized from the outset,
provide  an alternative computational fast tool to study   dynamical
processes in spatially distributed systems. As the coupling with nearest-neighbours
mimics the diffusion process, CML may be reckoned as a non-linear version of the diffusion process
on a lattice.  As a matter of fact, it is the  subtle  balance between
diffusion  and  nonlinear effects  that
leads   to the  patterns and non-trivial time behaviors found in CML.

\noindent
Pure diffusion on lattices has a long and venerable tradition in
statistical mechanics \cite{Haus87}. It was originally motivated
by solid-state applications as toy models for particle transport
in ordered and disordered solids, but it has soon grown into a field in its own right.
Interesting issues arise  when a quenched disorder is introduced to
model the fact that real life diffusion
occurs in disordered media such as a porous medium (see e.g. \cite{Haus87} for
a review). The presence of disorder drastically modifies
the diffusion properties, sometime in a somewhat unexpected way
\cite{Giacometti96}.
Depending on its type and its strength,
the disorder either inhibits diffusion (subdiffusion process) or
enhances it by creating priviledged paths (superdiffusion).

\noindent Motivated by this scenario, the natural question  arises of what would happen for CML in the presence
of a quenched disorder, where  three different mechanisms (non-linearities, diffusion and disorder)
compete against one-another. This  issue has already been addressed
for various non-linear systems \cite{Radons04}. 
Stochastically induced synchronization driven by the presence of
additive noise has also been observed in spatially extended systems \cite{Baroni01}.

Clearly the presence of disorder induces
a translational invariance breaking which generates a particular case
of  a system with a large number of
degrees of freedom which are non equivalent. In analogy to what is
observed  in other problems of  statistical physics, new phenomena may arise. As a matter of fact,
aging, dynamical phase transition, anomalous transport have
been observed in  one-dimensional maps  \cite{Radons04}.

\noindent In the present paper, we focus on two-dimensional CML. Two such
 systems are investigated. On
the regular lattice, the first one displays  phase transitions (``ferromagnetic
ordering'') and, the second one, a non-trivial global dynamics
Disorder is introduced at the simplest
possible level, namely a prescribed fraction
of bonds are randomly disconnected. This   affects the  phase diagram of the system~:
even a tiny fraction of disconnected bonds markedly modifies
or destroys altogether, the dynamics found in the ordered lattice
case.

\noindent The paper is organized as follows. In Sect. \ref{sect2},
we briefly review some results on  diffusion on a lattice and
nonlinear coupled maps, while Sect.\ref{sect3} introduces
the way disorder is implemented in the lattice. Sect.\ref{sect4}
details the numerical results for both CML models studied
 and Sect.\ref{sect5} is devoted to some concluding remarks.

\section{Diffusion and coupled maps on a lattice}
\label{sect2}

The most general master equation governing the linear diffusion of particles on a lattice reads
\begin{eqnarray}
\partial_t P_{x_ox}(t) &=& \sum_{y(x)}
\left[ W_{xy} P_{x_0y}(t) - W_{yx} P_{x_0x}(t) \right]
\label{eq1}
\end{eqnarray}
where $P_{x_ox}(t)$ stands for the probability of being at site $x \in {\cal Z}^d$
of a hypercubic lattice of dimension $d$, having started at site $x_0$ at the initial
time $t_0=0$. $W_{xy}$ denotes the transition rate
for a jump $y \to x$ and is  different from zero only for   nearest-neighbours $y$ of site $x$ (the notation $y(x)$
in the sum reflects this constraint).
A time discretized version (in time units such that $\Delta t=1$) reads
\begin{eqnarray}
P_{x_ox}(t+1) &=& (1-\epsilon_x) P_{x_ox}(t)+
\sum_{y(x)} W_{xy} P_{x_0y}(t)~~\hbox{where}~~\epsilon_x = \sum_{y(x)} W_{yx}
\label{eq3}
\end{eqnarray}
If only  positive symmetric transition rates  $0 \le W_{xy}=W_{yx}$   are allowed
and   $\sum_{y(x)} W_{xy} = 1$ (for all $x$), then one easily   verifies
that (a) $ 0 \le P_{x_ox}(t) \le 1$ and (b)  the total probability is conserved. The physical process associated
to such a mathematical framework can be described as  follows~:   at
each time $t$ a particle at $x$ can either
   move to a nearest-neighbour $y(x)$ with rate $W_{yx}$ or stay at
   $x$ with probability $1-\epsilon_x$. 
Quantities  $W_{xy}$ are  generally random
with the {\it proviso}
that condition $0 \le \epsilon_x \le 1 $ should hold true for any $x$.
Standard diffusion  corresponds to $W_{xy}=1/z$ where $z=2d$ denotes
the coordination number of a d-dimensional  hypercubic lattice.
Introducing  a quenched disorder in the
transition rates $W_{xy}$  changes the classical diffusion behavior into a sub-diffusive or super-diffusive 
one~\cite{Giacometti96}. Such a toy model may be used to understand   a variety of complex systems such as optical
excitation, polymers and diffusion-limited binary reaction~\cite{Haus87}.

\noindent Other   interesting dynamical
features arise  when non-linear terms are introduced in the local dynamics.
Let $C_{x}(t)$ be a real variable defined at point $x$ and time
$t$. This quantity belongs  to the interval $[-1,1]$ and
satisfies
\begin{eqnarray}
C_{x}(t+1)&=& (1-zg) F\left[C_{x}(t)\right]+ g
\sum_{y(x)}  F\left[C_{y}(t)\right]
\label{eq9}
\end{eqnarray}
in which   $F(C)$ stands for a non-linear homorphism  in  $[-1,1]$.
Quantity $C_{x}(t)$   may represent  the concentration
of a binary mixture, the values $-1$ and $+1$ corresponding to a
pure phase of the first or second  component respectively.
The above equation   resembles   equation (\ref{eq3}) with
 $W_{xy}=g$ for all nearest-neighbours but with a non-linear  map $F(C)$.
Coupling $g$ between neighbour sites   is constrained to satisfy $0\le g \le 1/2d$ so that, if the
initial concentration is in the range $[-1,1]$, it remains within the same interval for all times.
The first r.h.s factor  in equation  (\ref{eq9}) accounts for the
``self-interaction'' term, that is the coupling of each site with itself.
As $g \to 0^{+}$,  each site evolves
independently of the other (``high temperature regime''). At
$g = 1/(2d+1)$,  each site is coupled  with all its nearest-neighbours as well as with
itself  in an identical manner (``democratic coupling''). Finally as $g \to g_{max}=1/2d $, the self-interaction term vanishes
and all nearest-neighbour sites are strongly and equally coupled (``low temperature regime'').
For a two-dimensional lattice, the democratic coupling and the low-temperature regime correspond
to $g=0.2$ and $g=0.25$ respectively.

\noindent  In the present work,  we consider a finite size lattice
consisting of $L$ sites along each spatial direction with standard periodic
boundary conditions.
Two   maps $F(C)$ are studied, each one representing  a particular class  of dynamics. The first one  is the
Miller and Huse map~\cite{Miller93a}~\cite{Lemaitre99}
\begin{eqnarray}
F_{\mu}\left(C\right) &=&
\left\{
\begin{array}{ll}
-2\mu/3-\mu C \qquad \qquad  \mbox{for } -1 < C < -1/3 \cr
\mu C \qquad \qquad  \mbox{for } -1/3 < C < 1/3 \cr
2\mu/3-\mu C  \qquad \qquad  \mbox{for }  1/3 < C < 1
\end{array}
\right.
\label{eq14}
\end{eqnarray}
where $C \in [-1,1]$. The map (\ref{eq14}) with $\mu=3$ was introduced by Miller and Huse
to mimic  a statistical  system having two equilibrium
states i.e. a Ising-like dynamics. This model is appropriate to describe
a spinodal decomposition mechanism if variable $\sigma_x(t)  \equiv \textrm{sign} \left[C_{x}(t)\right]  =\pm 1$
plays the role of an Ising spin. In that case, a magnetization can be  defined as follows:
\begin{eqnarray}
m(t)&=&\frac{1}{V} \sum_{x} \sigma_x(t)
\label{eq12b}
\end{eqnarray}
where $V=L^d$ is the total number of sites.

\noindent The second study  concerns
the  logistic map
\begin{eqnarray}
F_{\mu}\left(C\right)&=&  1-\mu C^{2} \qquad 0 \le \mu \le 2  \qquad C \in [-1,1].
\label{eq11}
\end{eqnarray}
This CML   possesses a number of interesting dynamical features~\cite{Lemaitre99}. In particular, 
the spatially averaged quantity
 \begin{eqnarray}
\overline{C(t)} &=& \frac{1}{V} \sum_x C_x(t)
\label{eq17}
\end{eqnarray}
possesses a macroscopic time-periodic  evolution
 while spatial patterns similar to those found for Miller-Huse map  are not visible. 
 Moreover the graph of the asymptotic time evolution of $\overline{C(t)} $, with respect to parameter $\mu$
displays a bifurcation diagram similar to the one observed for a unique Logistic map \cite{Lemaitre98}.

\section{Coupled Map Lattice and Disorder.}
\label{sect3}

\noindent  In close analogy with  linear diffusion   on a lattice,
the effect of a quenched disorder is introduced on the  two CML
 discussed above. The generalization of Eq.(\ref{eq9}) then reads
\begin{eqnarray}
C_{x}(t+1) &=&  (1-\epsilon_x) F\left[C_{x}(t)\right]+
\sum_{y(x)} W_{xy} F\left[C_{y}(t)\right]
\label{eq13}
\end{eqnarray}
Paralleling the diffusion case,   couplings  are partially random and verify for arbitrary~$x$
\begin{eqnarray}
W_{yx}=  W_{xy},~~~0<\epsilon_x &\equiv& \sum_{y(x)} W_{yx} \le 1.
\label{eq13a}
\end{eqnarray}
As a consequence, quantities $ C_{x}(t)$ always remain  within the original interval $ [-1,1]$ since
 \begin{eqnarray}
|C_{x}(t+1)| \le  (1-\epsilon_x)  + \sum_{y(x)} W_{xy} \le  (1-\epsilon_x)  + \sum_{y(x)} W_{yx} =1.
\label{eq13c}
\end{eqnarray}
This model includes  all the previous cases:
\begin{description}
\item{a)} Linear diffusion   with disorder (\ref{eq3})  when $F(x) \equiv x$.
\item{b)} Nonlinear coupled maps (\ref{eq9})  when $W_{xy}=g$  (and hence $\epsilon_x = z g$).
\end{description}
\noindent In the following, a  quenched binary disorder is used satisfying  the   distribution
\begin{eqnarray}
\rho(w)&=& (1-p) ~ \delta(w-g)+p \delta(w)
\end{eqnarray}
where $p$ stands for the percentage of connections which are cut. The value $p=0$ corresponds
to the  regular case. It should be emphasized that, as bonds  are removed, the
self-interaction term increases~(see Eq.(\ref{eq13})).
Disorder may set a competitive mechanism disfavouring the balance involving  couplings between
lattice points, which
tends to "synchronize" the maps spatially or temporally
and the non-linearity of the maps, which tend
to destabilize them. Disorder
is thus expected to act  as a further destabilization  preventing the onset of an "ordered" phase.

\section{Numerical results}
\label{sect4}

\noindent Numerical results are obtained   on a square lattice ($d=2$).
The  Miller and Huse case is first presented  in Sect. A,B,C as a
prototype of a spinodal decomposition  mechanism. Sect. D and E are devoted to the logistic map
where a global   time evolution is observed.

\subsection{The ordered Miller and Huse phase diagram for $\mu=3$ and $\mu=1.9$}

\begin{figure}[hbtp]
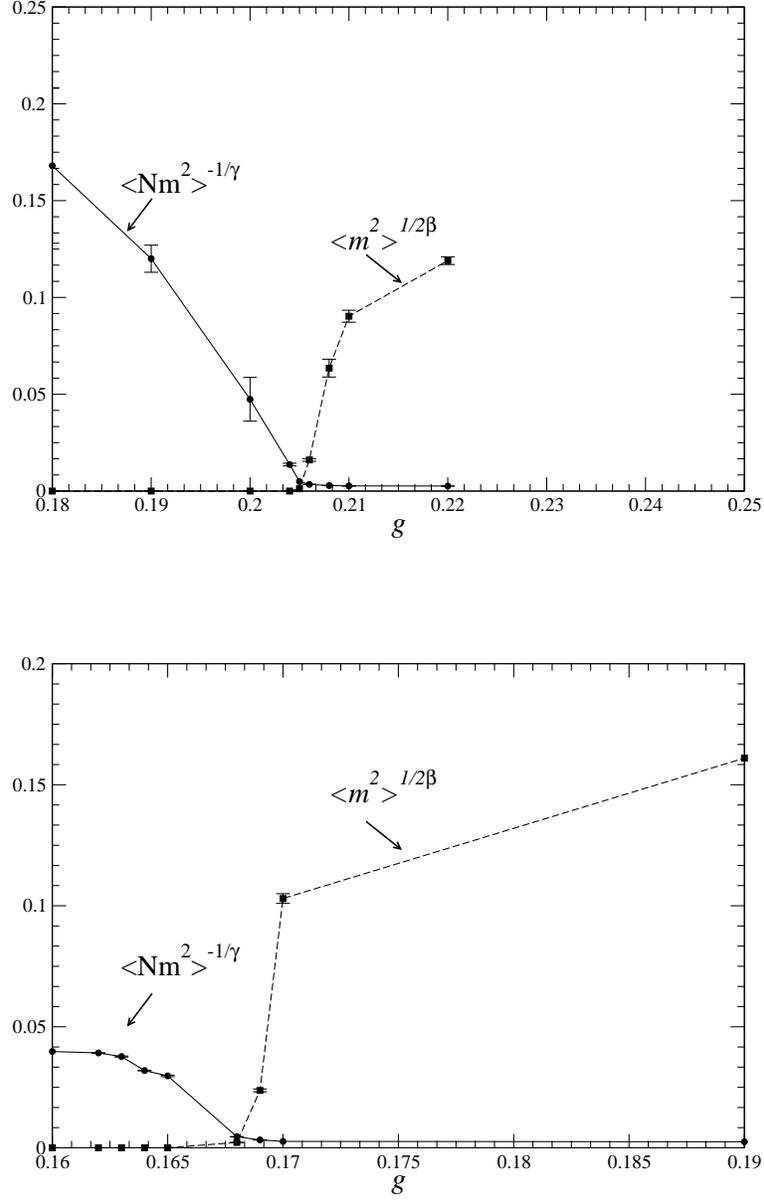

\begin{center}
\includegraphics[width=10cm]{Fig1a.eps}\\
\vskip1.5cm
\includegraphics[width=10cm]{Fig1b.eps}
\end{center}
\caption[]{The phase diagram for the Miller and Huse map for $\mu=3$ (top) and
$\mu=1.9$ (bottom). The  lattice sites are $N=L\times L$ with  $L=128$ and magnetization is computed once
$T=2 \times 10^5$ iterations have been performed and
exponents $\gamma$ and $\beta$ are those identified in \cite{Marq97}.
\label{fig1}
}
\end{figure}

\noindent Fig.\ref{fig1} (top) depicts  the Miller-Huse  phase diagram  for $\mu=3$
\cite{Miller93a}. If $m^2 $ denotes the  value
of the mean-square magnetization, this quantity is raised to two suitable exponents
$\left \langle m^2 \right \rangle^{\frac{1}{2\beta}}$ and $ \left \langle L^2m^2 \right \rangle^{-
\frac{1}{\gamma}}$, and plotted as a function of the coupling strength $g$.
In their original paper, Miller and Huse used the  two dimensional Ising critical exponents
$\beta_{Ising}=1/8$ and $\gamma_{Ising}=7/4$ \cite{Stanley71}. A successive
and more refined work~\cite{Marq97} identified the characteristics exponents
values $\beta \sim 0.111$, $\gamma \sim 1.55$  which are close but not
identical to those of the two dimensional  Ising model.
Figure \ref{fig1}  was obtained for lattice size $L=128$ with a  magnetization which was computed after
$T=2 \times 10^5$ iterations which appear to be large enough to reach a steady state, and
averaged over $10-50$ different initial conditions ($50$ refers to the left branch
of the phase diagram $g \le g_c$, where fluctuations are much higher).
As one approaches the critical value $g_c = 0.20(5)$, a phase transition occurs.
This result is  in  agreement with those of~\cite{Marq97} which pinned down the
transition value at $g_c = 0.2051(5)$ on the basis of  more
extensive simulations. Fig.\ref{fig1} (bottom) provides the same
computations for  $\mu=1.9$ and indicates a critical value $g_c =
0.16(8)$   not previously reported. Here and below the parenthesis
indicates the degree of uncertainty of the result.

\begin{figure}[hbtp]
\begin{center}
\includegraphics[angle=270,width=8cm]{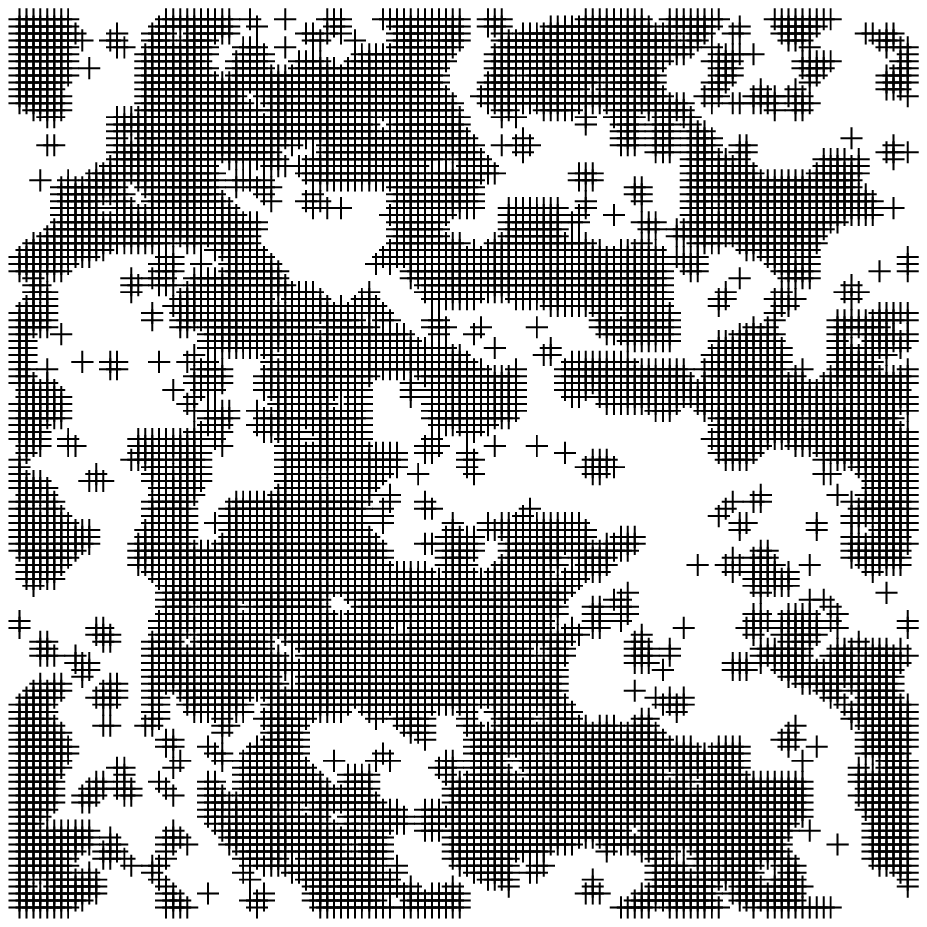}
\includegraphics[angle=270,width=8cm]{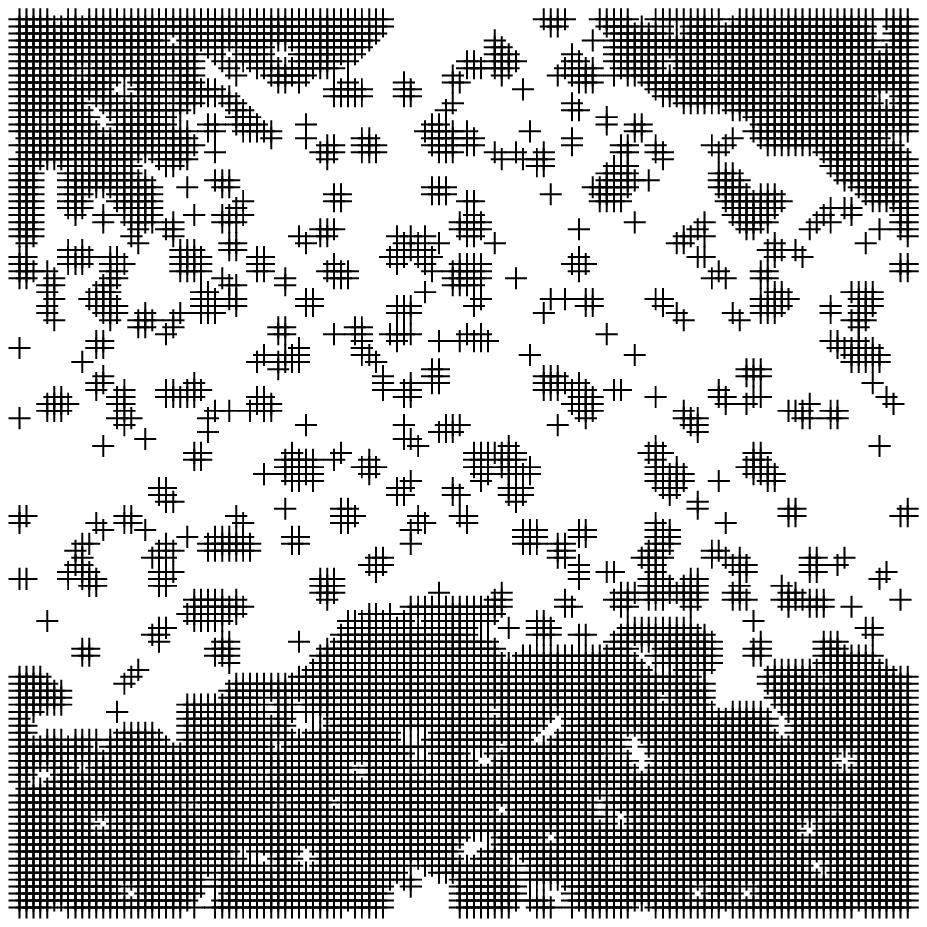}\\
\vskip1.5cm
\includegraphics[angle=270,width=8cm]{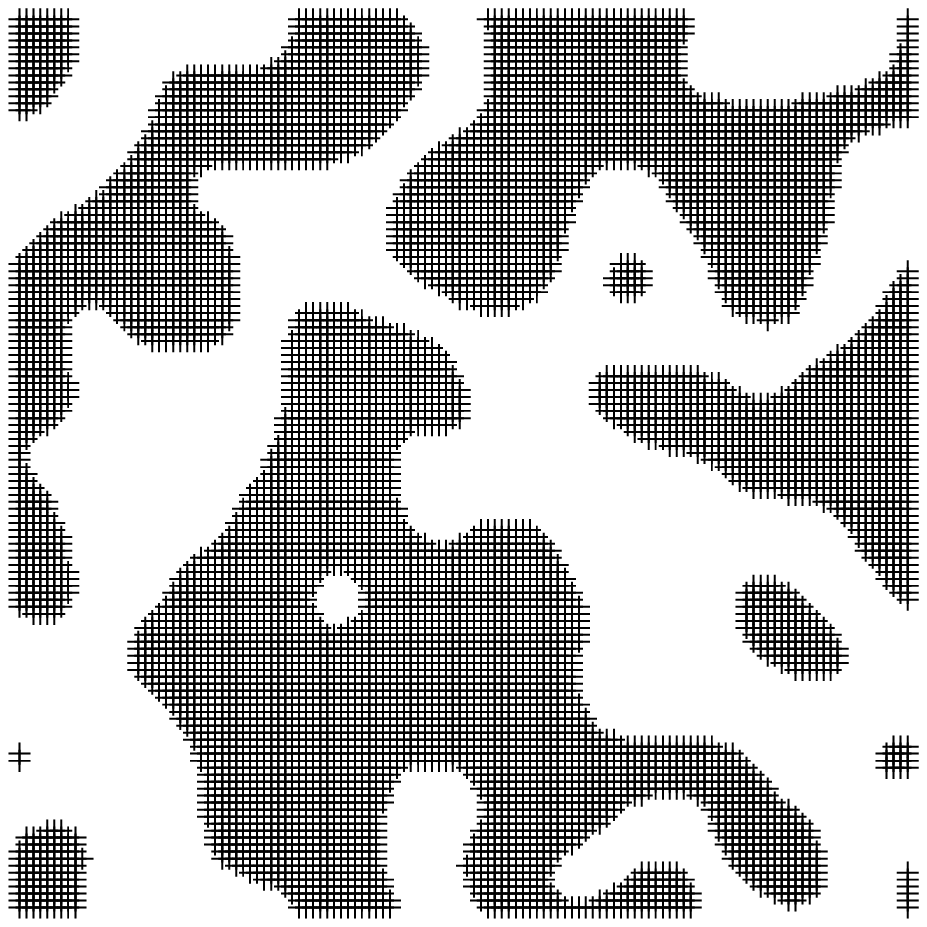}
\includegraphics[angle=270,width=8cm]{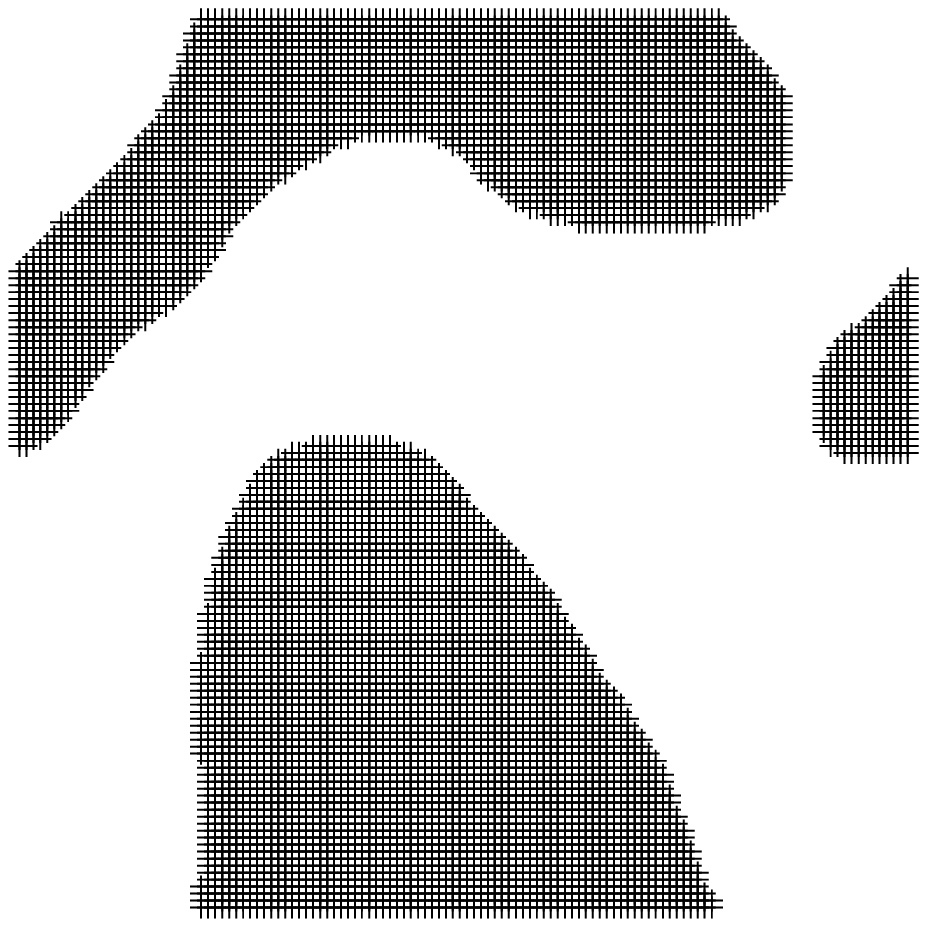}
\end{center}
\caption[]{Snapshots for the  Miller and Huse CML on the regular lattice for $g > g_c$.
Case $\mu=3$, $g=0.21$ (top) with $t=100$~(left), $t=1000$~(right).
Case $\mu=1.9$, $g=0.20$ (bottom) with $t=100$  (left),  $t=1000$  (right).
The  lattice size is $L=128$.
\label{fig2}
}
\end{figure}

\noindent The fact that the total magnetization is not conserved,
suggests that this  dynamics could be akin to a dynamical Ising model
with non-conserved magnetization (model A)\cite{HohenbergHalperin,Krug}.
However, the critical
coupling, as well as
the time evolution~(see below) $g_c$ changes for different $\mu$. This feature is apparent in
Fig. \ref{fig2} where, starting from a random and uncorrelated initial
condition for $g > g_c(\mu)$,   two snapshots are depicted  for $\mu=3$
(top) and $\mu=1.9$ (bottom) at  stages  $t=100$~(left) and $t=1000$~(right).

\subsection{The persistence probability in the case of a regular lattice}

\begin{figure}[hbtp]
\begin{center}
\includegraphics[width=10cm]{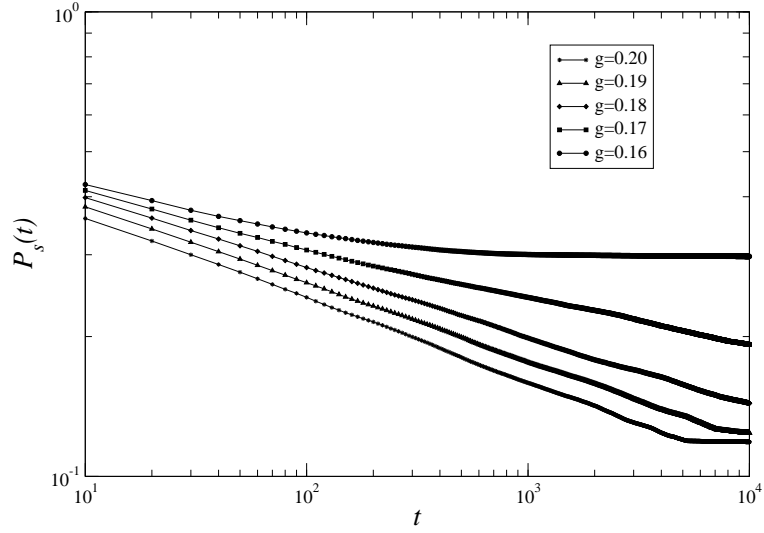}
\end{center}
\caption[]{The persistence probability for the Miller and Huse maps on
  the regular lattice
for $\mu=1.9$ ($g=0.16$ is the uppermost curve). The  lattice size is $L=128$.
\label{fig3}
}
\end{figure}

\noindent Another interesting quantity, is the
 persistence probability $P_s(t)$, which measures the fraction of
 spins which have not changed sign up to time $t$. This concept
was first introduced  within this context in~\cite{Lemaitre99}.
It was  shown that, for  $\mu=1.9$, a
transition occurs at  $g_e\sim 0.169$  appearing as an
abrupt modification in the slope of   probability $P_s(t)$~(Fig \ref{fig3}).
For $g_e \le g $,  this probability takes the form
\begin{eqnarray}
P_s(t) &\sim& t^{-\theta}~~~\hbox{with }~~\theta(g) \sim (g-g_e)^{w}
\label{eq15}
\end{eqnarray}
Values $w \sim 0.20$ and $g_e = 0.16(9)$ are reported in~\cite{Lemaitre99}. This latter number
is strikingly close to the critical value $g_c = 0.16(8)$ found in the above section.
The transition of the
persistence probability thus appears to be  a  consequence of the phase transition of
magnetization  (and not computed in~\cite{Lemaitre99}), rather than  a different transition.

\subsection{ Miller-Huse coupled map lattice with binary quenched disorder}

\begin{figure}[hbtp]
\begin{center}
\includegraphics[width=10cm]{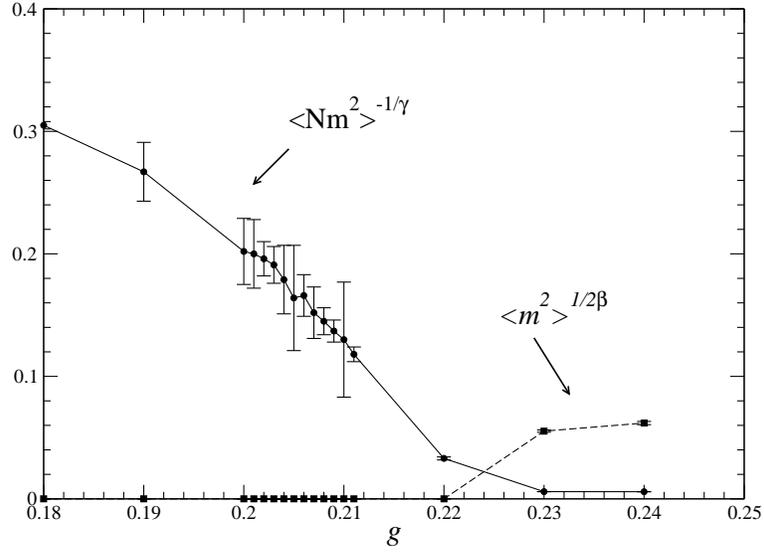}\\
\vskip1.5cm
\includegraphics[width=10cm]{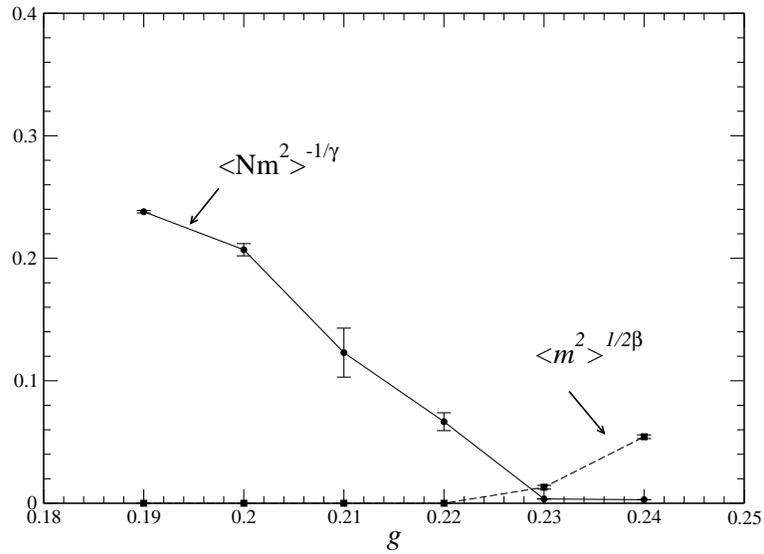}
\end{center}
\caption[]{Phase diagram for Miller and Huse maps for $\mu=3$
  and dilution: $p=0.05$ (top) and $p=0.065$ (bottom).
Lattice size is $L=128$. Magnetization is computed once
$T=2 \times 10^5$ iterations have been performed and
exponents $\gamma$ and $\beta$ are those identified in \cite{Marq97}.
\label{fig4}
}
\end{figure}

\begin{figure}[hbtp]
\begin{center}
\includegraphics[width=10cm]{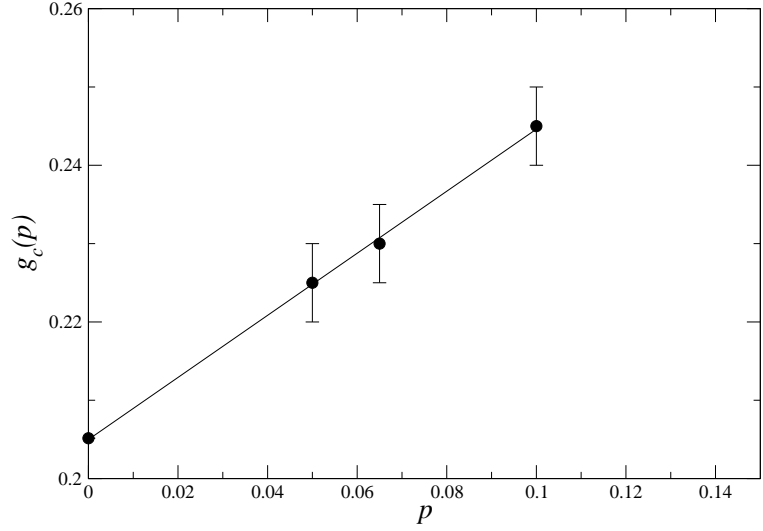}
\end{center}
\caption[]{Critical coupling $g_c$ as a function of  the dilution
  parameter $p$. Miller-Huse  model  for $\mu=3$. The solid line
represents a best fit result yielding $p_c \sim 0.11$ for $g_c =0.25$.
\label{fig5}
}
\end{figure}

\noindent Let us now introduce a slight disorder on the
Miller and Huse coupled maps. Two  small percentages of disconnected bonds have been used~: $p=0.05$ and $p=0.065$.
Results for $\mu=3$ are reported in Fig.~\ref{fig4}. Note that
exponents of the undiluted case are employed on the basis
that critical exponents of the Ising model with random binary coupling
 are found  to be similar  to those computed for a regular lattice~\cite{Roder99}.
Dilution  has the effect of shifting the critical coupling $g_c$ to higher values.
This shift can be qualitatively understood by a simple argument. Let
denote $z(p)=z(1-p)$ the average effective coordination number with dilution $p$. Within
a mean field description, the average coupling strength   increases
in the presence of dilution by a factor
\begin{eqnarray}
\frac{g_c\left(p\right)}{g_c\left(p=0\right)}&=& \frac{1}{1-p}
\end{eqnarray}
thus leading to a linear dependence for $g_c(p)$ for small values of $p$.
Our calculations reproduce  this linear behavior
(Fig. \ref{fig5}), although with a different numerical value of slope.

\begin{figure}[hbtp]
\begin{center}
\includegraphics[width=10cm]{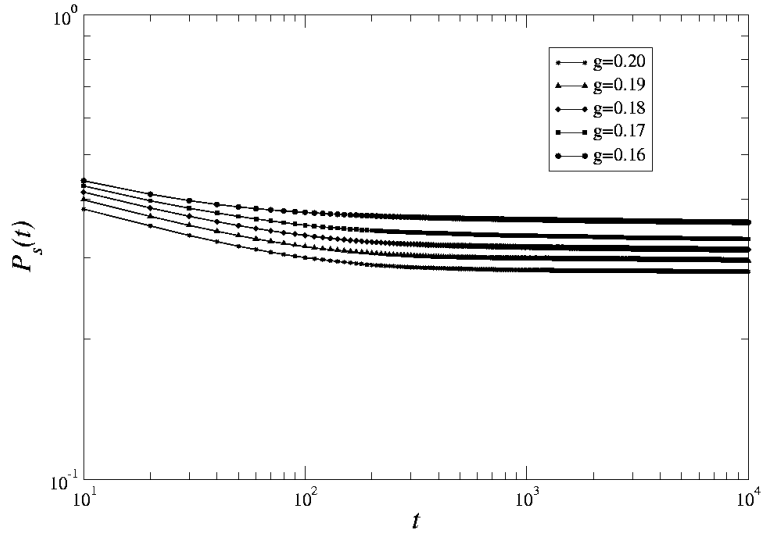}
\end{center}
\caption[]{Persistence for the Miller and Huse map for $\mu=1.9$ and dilution $p=0.05$.
\label{fig6}
}
\end{figure}

\noindent  As the maximum possible  $g$ is equal  to a constant
$g_{\textrm{max}}=1/2d=0.25$,
the right branch of the phase diagram shrinks as $p$ increases, and
there will be a critical threshold $p_c$, such that $g(p_c)=g_{\textrm{max}}$,
above which a phase transition is no longer observed.
Results of Fig.\ref{fig5} then suggest $p_c \sim 0.11$,
although a more extensive analysis would be needed for a precise estimate. We
checked that, for values $p>p_c$, the ferromagnetic transition is
indeed not observed.
A small percentage of disconnected bonds
therefore disfavours a collective behavior and can even inhibit
the transition. A similar trend is mirrored by the persistence probability
(Fig. \ref{fig6}), which  is computed with a dilution $p=0.05$ and under the same conditions as in
Fig. \ref{fig3}. This also supports the   equivalence between persistence and
ferromagnetic transitions.

\begin{figure}[hbtp]
\begin{center}
\includegraphics[angle=270,width=10cm]{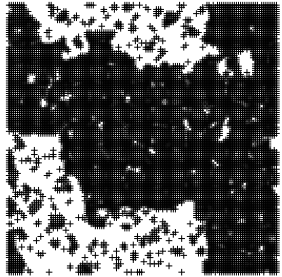}\\
\vskip1.5cm
\includegraphics[angle=270,width=10cm]{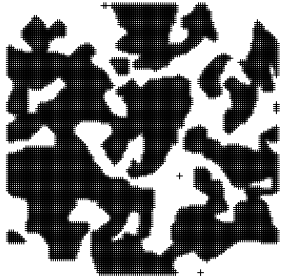}
\end{center}
\caption[]{Snapshots for the Miller and Huse map with dilution
  ($p=0.05$) for $g  > g_c(p)$ in
  the case $\mu=3$, $g=0.23$, $t=1000$ (top), and in the case $\mu=1.9$, $g=0.20$, $t=1000$  (bottom).
Here the size is $L=128$.
\label{fig7}
}
\end{figure}

\begin{figure}[hbtp]
\begin{center}
\includegraphics[angle=270,width=10cm]{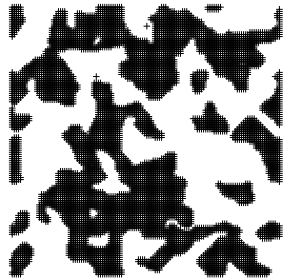}\\
\vskip1.0cm
\includegraphics[angle=270,width=10cm]{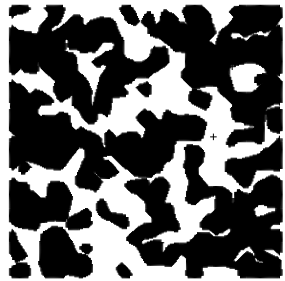} \\
\vskip1.0cm
\includegraphics[angle=270,width=10cm]{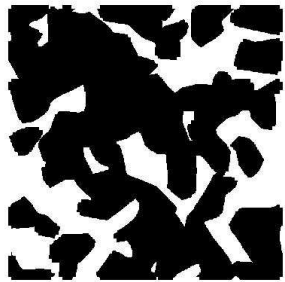}

\end{center}
\caption[]{ Various   rescaled patterns at different $p$ but constant $pL^2$.
Parameters are $g=0.2$ and $\mu=1.9$.
From top to bottom,  $L=128$ and $p=0.05$; $L=256$ and $p=0.0125$;  $L=512$ and $p=0.003125$.
\label{fig8}
}
\end{figure}

\noindent   Patterns for
both $\mu=3$ and $\mu=1.9$
above the critical coupling ($g > g_c(\mu)$) are  shown at some
intermediate time ($t=1000$) in Fig. \ref{fig7}
for dilution $p=0.05$. These are the analogous of those previously
obtained under the same conditions in the regular lattice (Fig. \ref{fig2}).
If patterns are pinned by defects, we expect similar patterns upon a
rescaling conserving the quantity $p
L^2$. Indeed, the number of  "defects" (sites with disconnected bonds)
should be equal to the fraction  $p$ of disconnected bonds times
the total surface  $N \equiv L^2$. Figure \ref{fig8} indicates that this is
precisely the case
as the patterns obtained at a relatively late
stage of the evolution ($t=10^4$)   $L=128,p=0.05$, and
$L=256,p=0.0125$, $L=512,p=0.003125$  are quite similar.

\subsection{The Logistic CML  and non-trivial collective behavior}

\noindent In the Miller-Huse CML, the emphasis was
on  collective spatial  behavior in the sense of ordinary critical
phenomena. By contrast,  the logistic CML (\ref{eq11}) is an example of
a collective time evolution~\cite{Lemaitre99}. Depending on  parameter
$\mu \in [0,2]$, a large variety of complex behaviors are observed.
Building upon previous analyses on this system with no disorder
\cite{Lemaitre99}, an extensive analysis is performed both with and without
disorder. Our results are in perfect agreement with
published results~\cite{Lemaitre99}~\cite{Lemaitre98}.
In addition, we address issues such as transient behavior, coupling dependence
and competition with disorder, that were not previously covered.

\begin{figure}[hbtp]
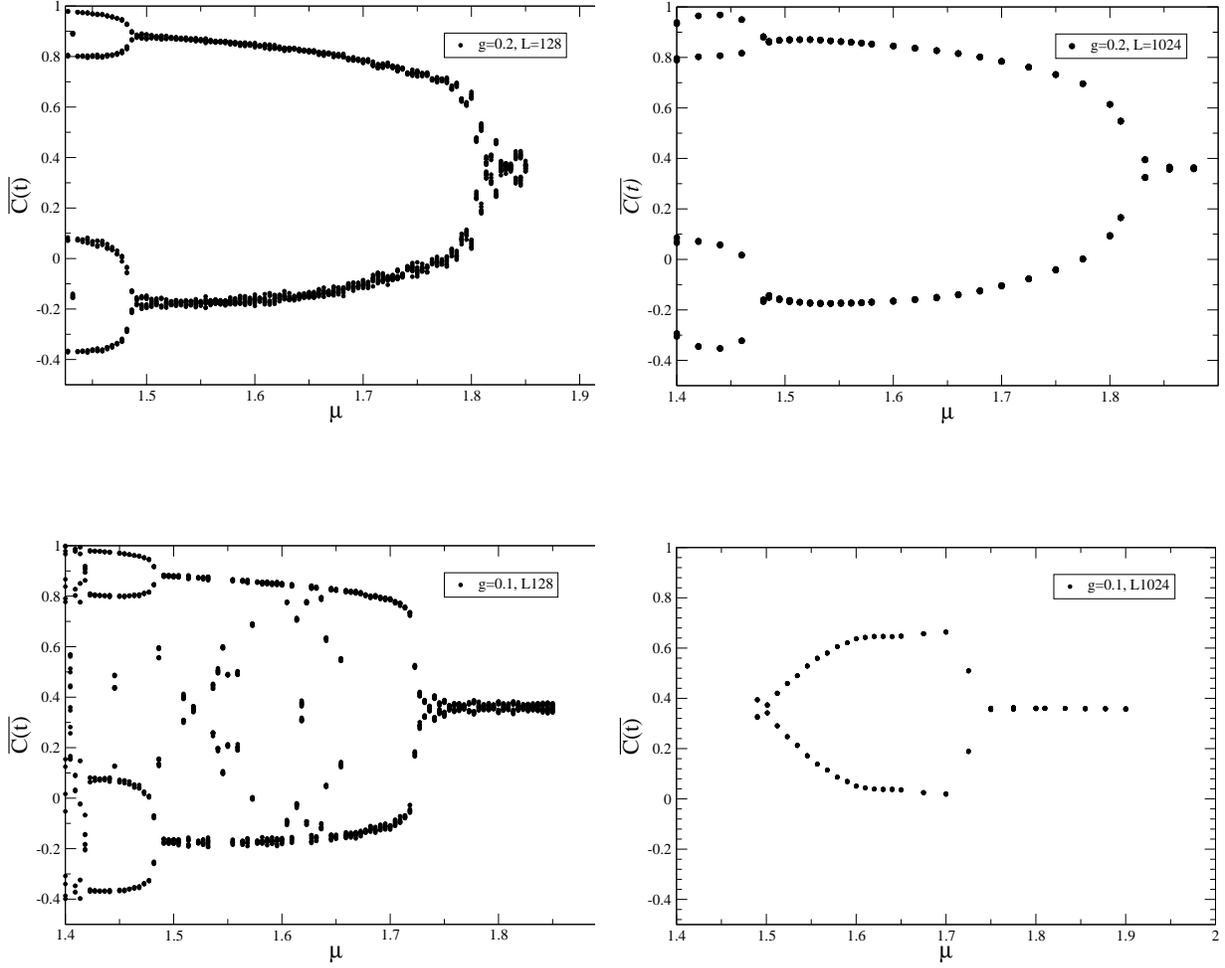

\begin{center}
\includegraphics[width=8cm]{Fig9a.eps}
\includegraphics[width=8cm]{Fig9b.eps}\\
\vskip1.5cm
\includegraphics[width=8cm]{Fig9c.eps}
\includegraphics[width=8cm]{Fig9d.eps}\\
\end{center}
\caption[]{ The   logistic CML on a regular lattice~:  the spatial average
$\overline{C(t)} = \frac{1}{V} \sum_x C_x(t)$ is shown as a function of  $\mu$
for the  six  consecutive times $T+1,T+2,T+3,T+4,T+5,T+6$ where $T=10^6$ for
$L=128$ and $T=10^4$ for $L=1024$.
System  sizes and couplings are $g=0.2$,  $L=128$ (top left),
$g=0.2$, $L=1024$ (top right), $g=0.1$, $L=128$ (bottom left), $g=0.1$, $L=1024$ (bottom right)
\label{fig9}
}
\end{figure}

\begin{figure}[hbtp]
\begin{center}
\includegraphics[angle=270,width=8cm]{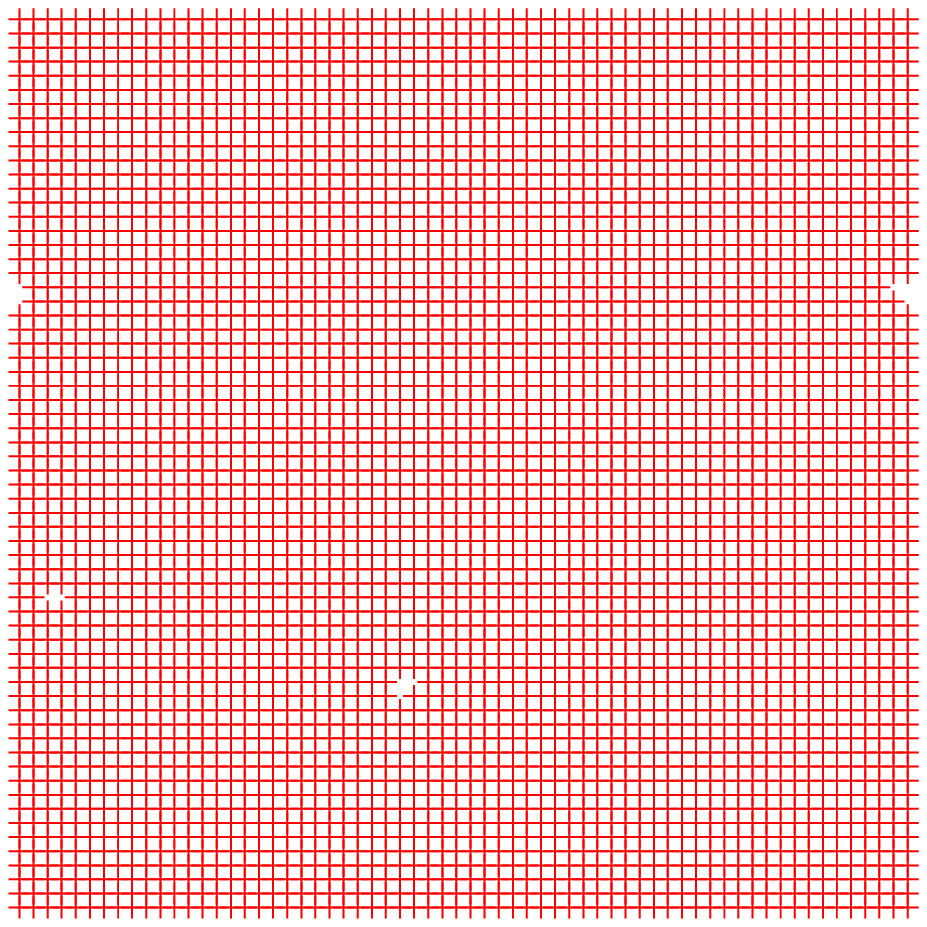}
\includegraphics[angle=270,width=8cm]{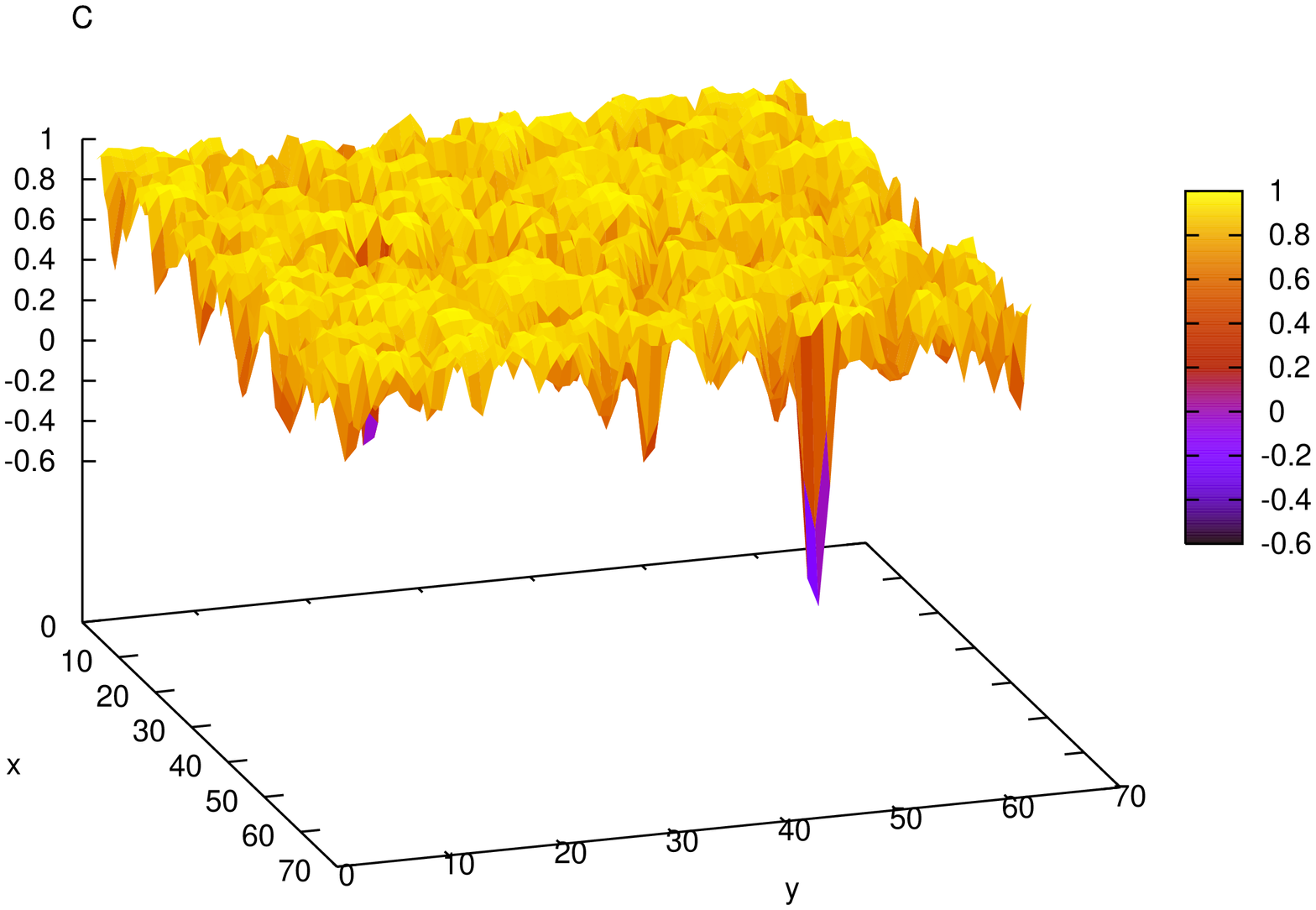}\\
\includegraphics[angle=270,width=8cm]{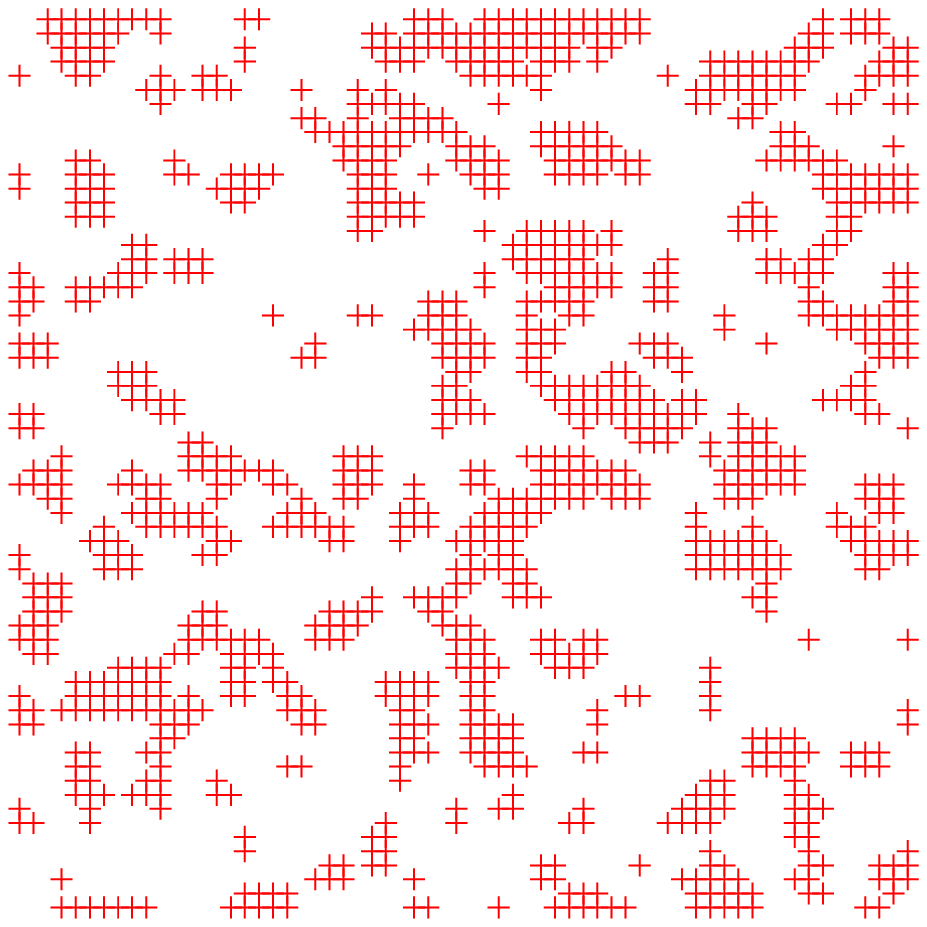}
\includegraphics[angle=270,width=8cm]{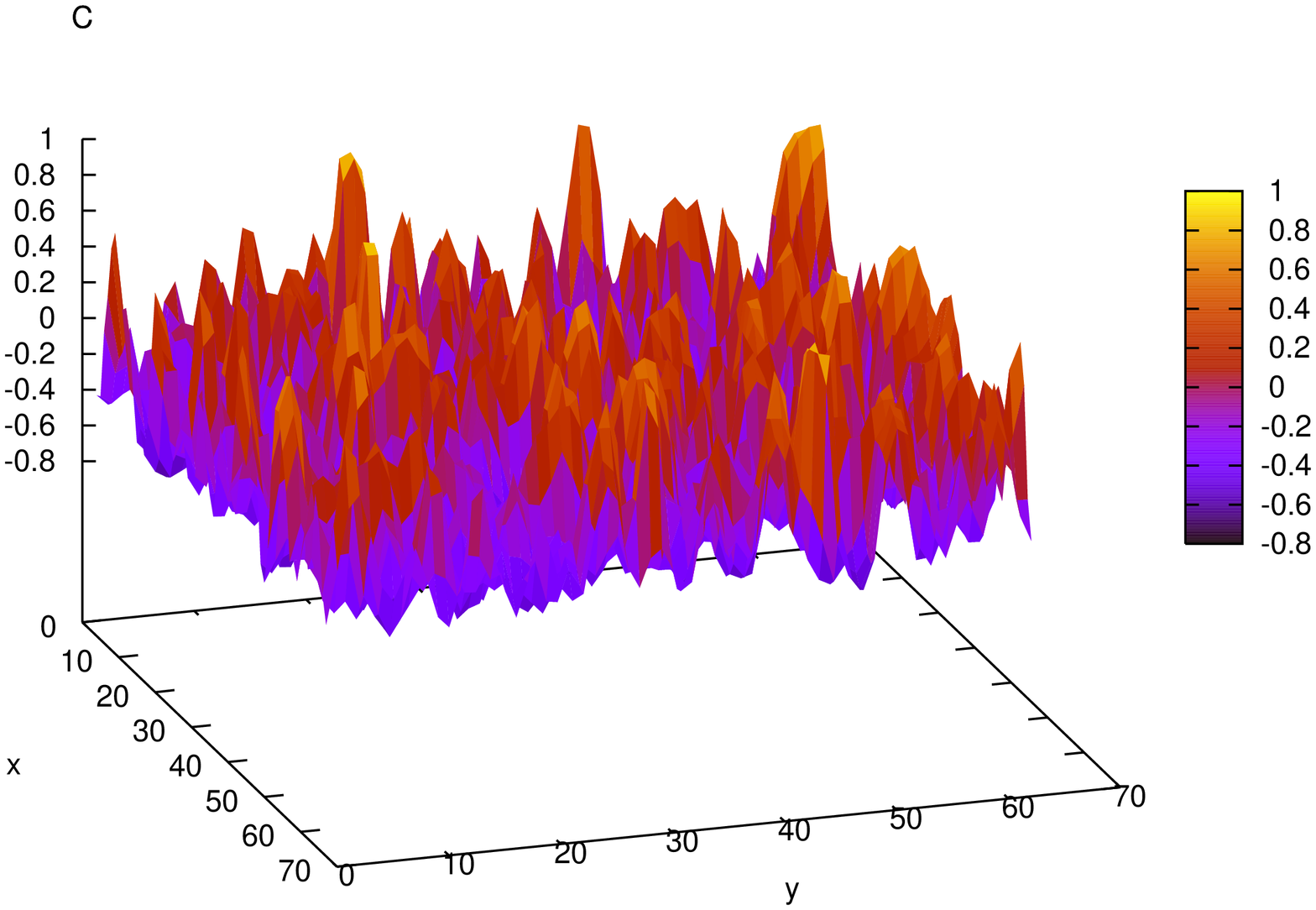}\\
\includegraphics[angle=270,width=8cm]{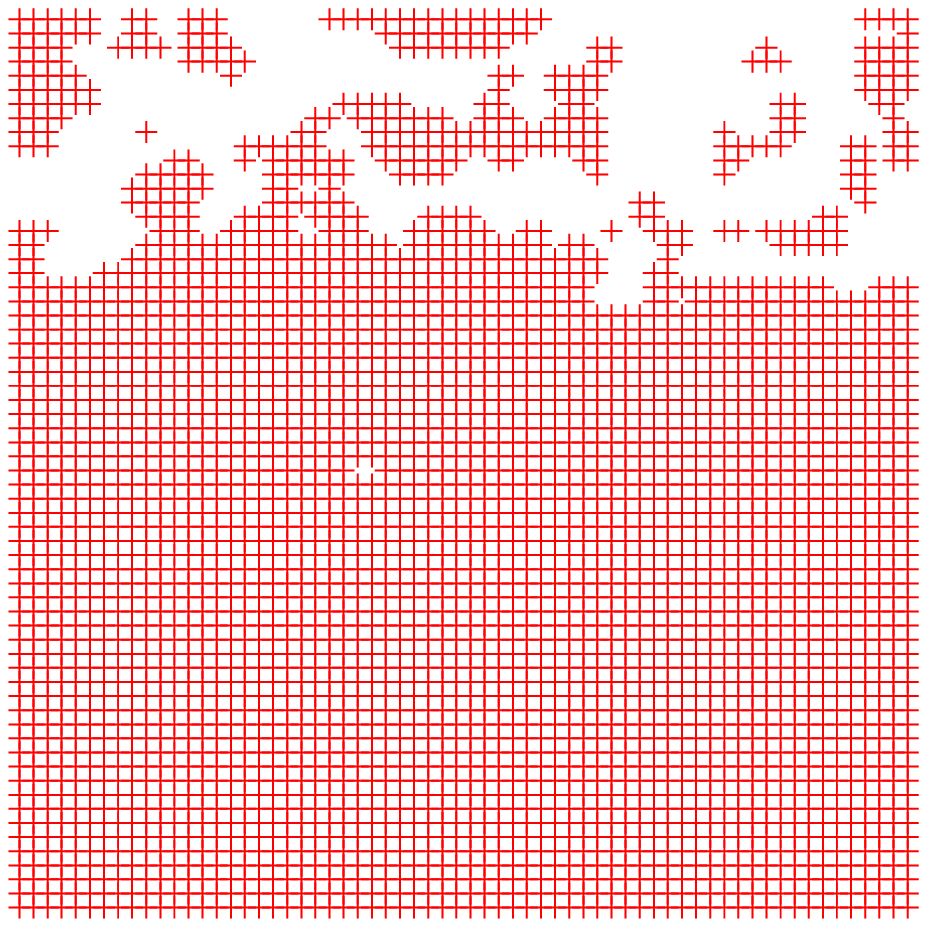}
\includegraphics[angle=270,width=8cm]{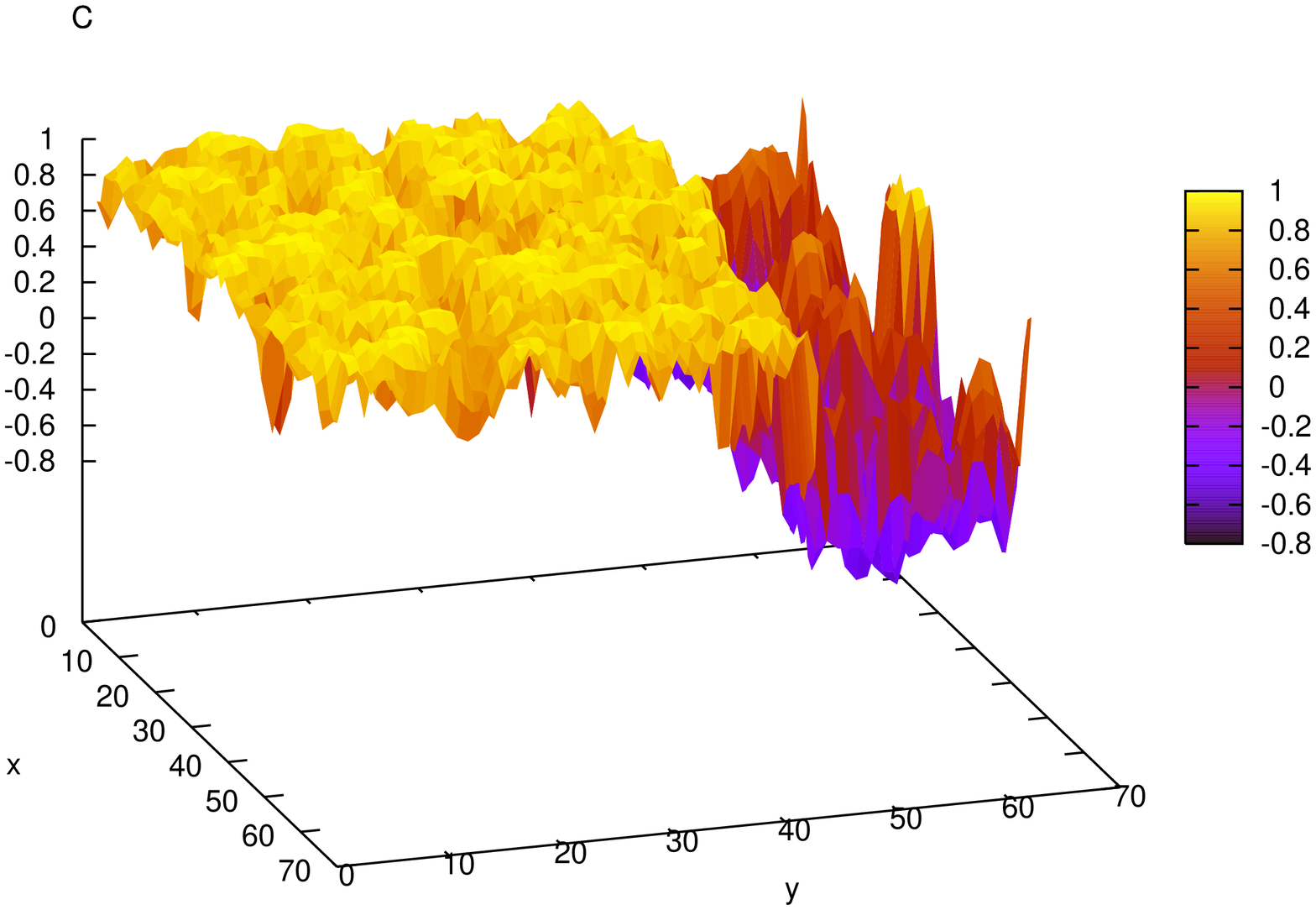}\\
\includegraphics[angle=270,width=8cm]{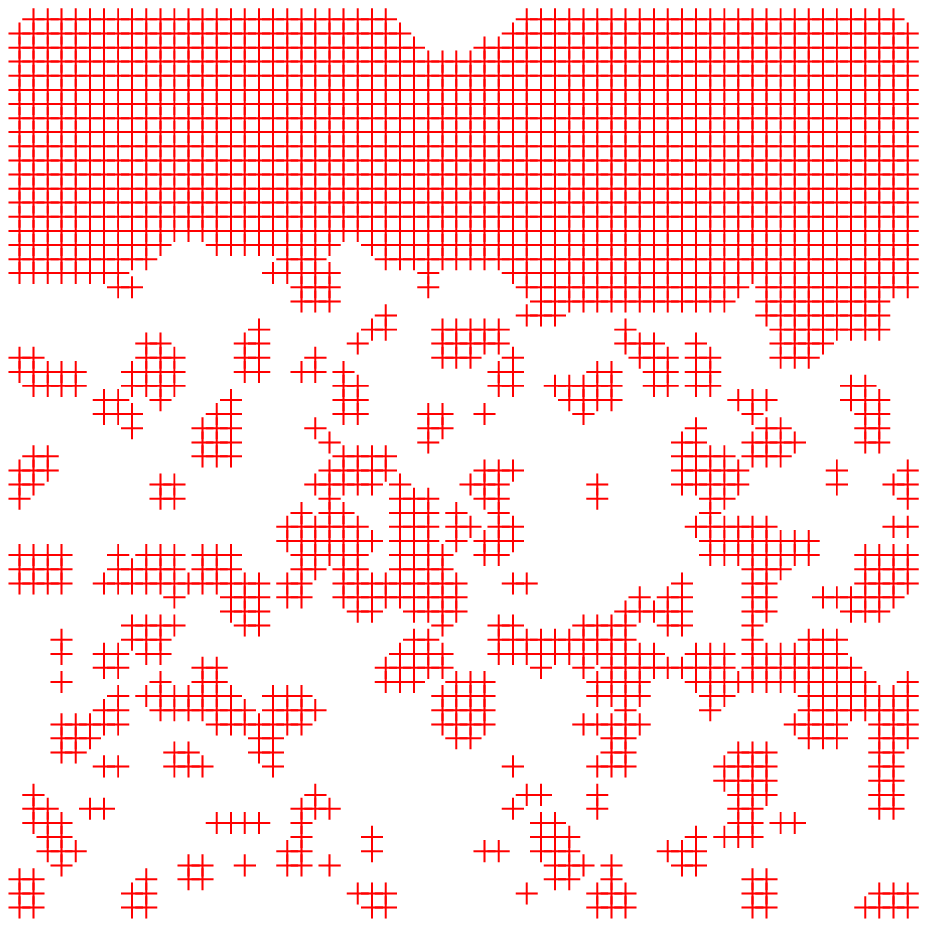}
\includegraphics[angle=270,width=8cm]{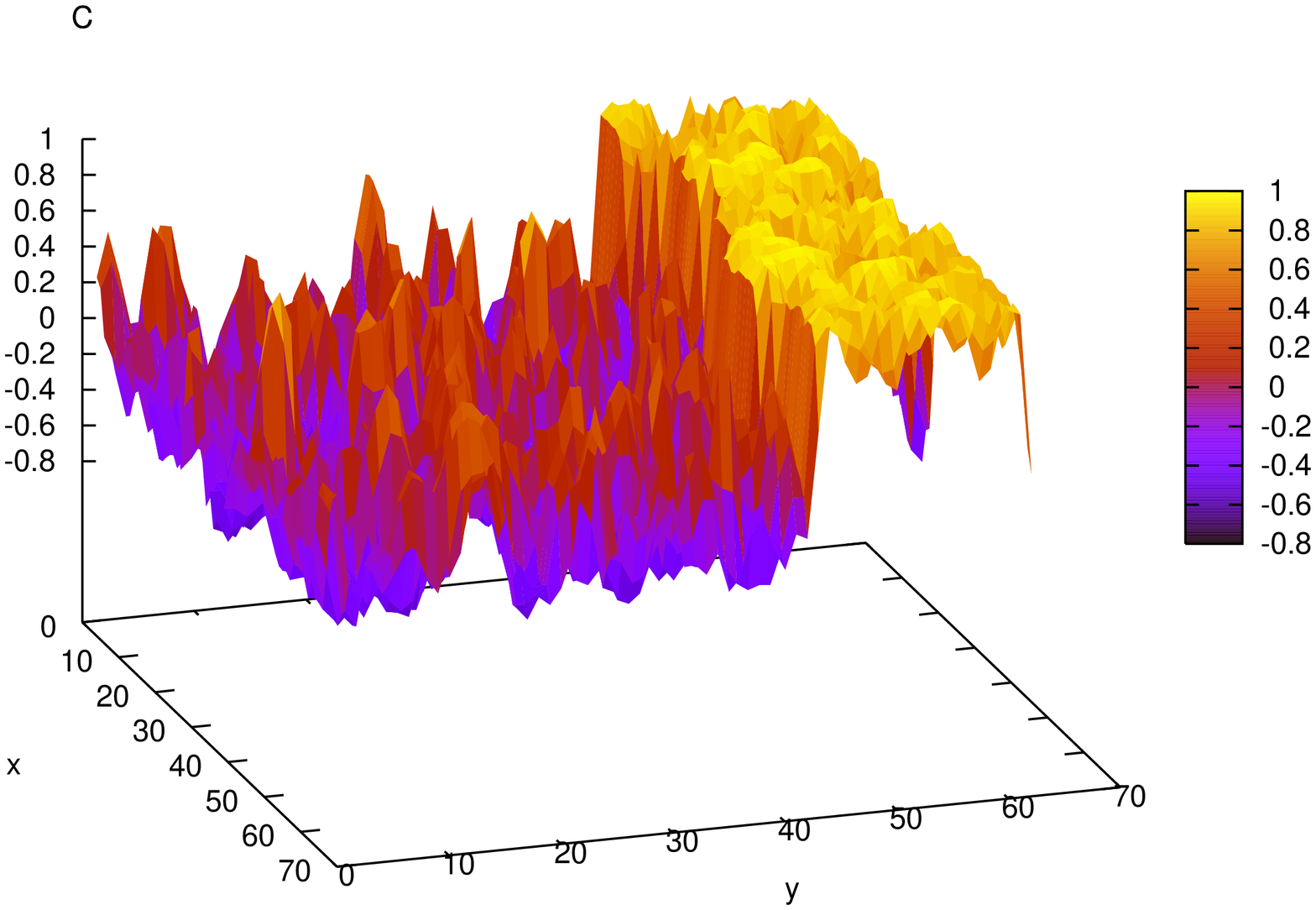}\\
\end{center}
\caption[]{(Color online) Snapshots of spin $\sigma_x(t)$ (left) and field  $C_x(t)$ (right)
for $\mu=1.66$, $g=0.1$, $L=64$, $T=10^6$. Two different random
initial conditions have been used for rows 1,2 and rows 3,4
respectively. Row 1 and 3 correspond to time $T$ and rows 2 and 4 to
time $T+1$.
\label{fig10}
}
\end{figure}

\begin{figure}[hbtp]
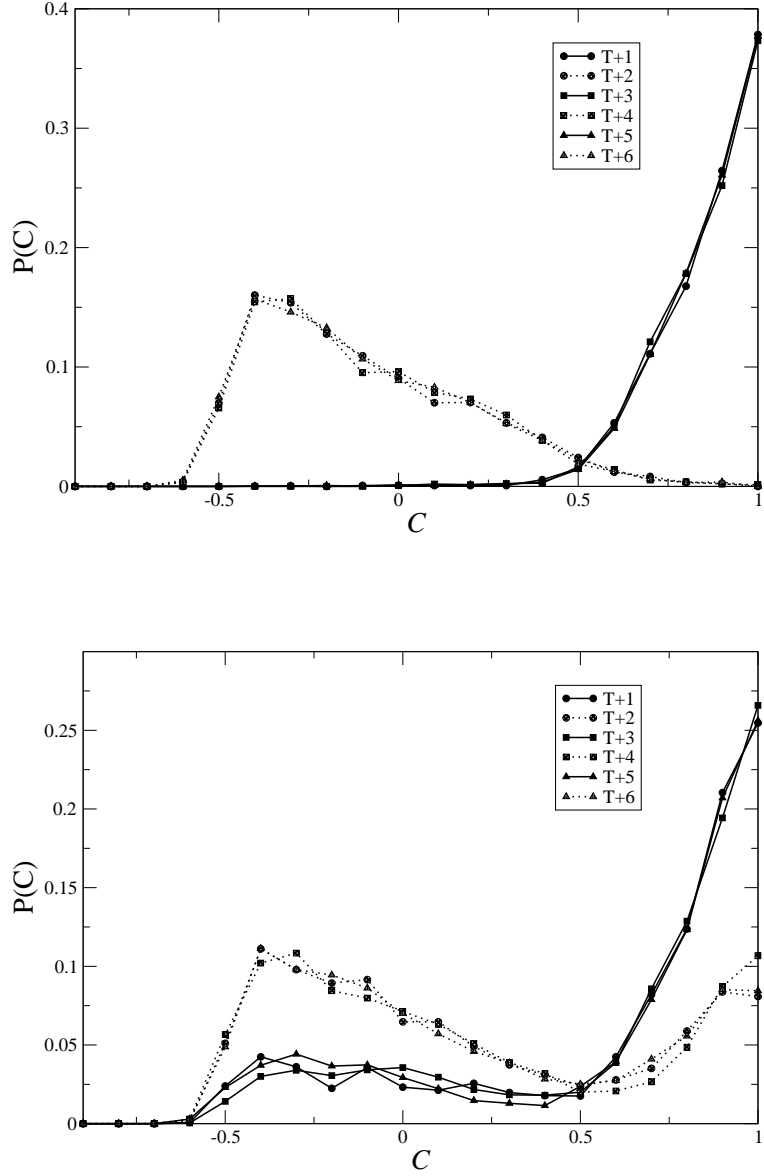

\begin{center}
\includegraphics[width=10cm]{Fig11a.eps}\\
\vskip1.5cm
\includegraphics[width=10cm]{Fig11b.eps}
\end{center}
\caption[]{Probability distribution function~(PDF)  of
variable $C(t)$ over the lattice
for the  logistic CML on a regular lattice ($p=0$) with $g=0.1$ and $\mu=1.66$. 
Computations are performed with a  size $L=64$ with top and bottom figures differing by the 
random initial condition.
Various  curves refer to six consecutives times $T+1,~T+2,~T+3,~T+4,~T+5,~T+6$
with $T = 10^6$.
\label{fig11}
}
\end{figure}

\noindent  System (\ref{eq9}) is first iterated for extremely long time
and  spatial average $\overline{C(t)}$ is thereafter computed   when a steady state
is expected to be attained. Note that some numerical tests require times up to $T=5
\times 10^6$ which would be prohibitively long for size $L=2048$. Hence
our investigations have been  performed with smaller
sizes ($L=1024$, $L=128$ and $L=64$). This steady state condition is checked for by
considering $\overline{C(t)}$ at several successive large times and
showing that the same sequence is  periodically found.
This sequence  is then analyzed as  a function of
parameter $\mu$ for different couplings $g$.
For  democratic coupling  $g=0.2$, a highly regular  time evolution arises
yielding a periodic doubling bifurcation  cascade  for
$\overline{C(t)}$. Fig.~\ref{fig9} reports results for
two computations: $L=128$ with $T=10^6$, and  $L=1024$ with $T=10^4$.
Both results nicely superimpose and are compatible with those of~\cite{Lemaitre99}~\cite{Lemaitre98} up to
fluctuations of order  $1/\sqrt{L}$ (in these work,a size  $L=2048$ was used).
This suggests that the CML behaves  in the thermodynamic
limit and that steady state has been reached in both cases.
For other couplings, the observed behavior can be
more involved. Indeed identical computations were repeated for
weaker couplings ($g=0.1$ in Fig.~\ref{fig9}).
For $L=128$, the same final time is used, and quite a few
"singular" points (i.e. outside the bifurcation diagram)
are now distinguishable. For size $L=1024$,
computations have been performed up to times as large as $10^4$ and
the system is always found in a transient state  (see last picture of
Fig.\ref{fig9}), in contrast with the $L=128$ case.
As in the uncoupled case $g=0$,  values of $C_x(t)$  are almost uniformly distributed with the exception of some
particular windows characteristic of the single map, it is then
plausible to expect longer transient
regimes and a stronger dependence on the initial conditions in the
weak coupling regime ($g<0.2$). Indeed Fig.~\ref{fig10} shows
  snapshots of  spins $\sigma_x(t)$  (left) and fields
$C_x(t)$ (right) at two extremely long times ($T$ and $T+1$ with $T=10^6$)
for  {\it identical parameters}~($L=64$, $g=0.1$, $\mu=1.66$). The two situations  only differs
by their   {\it  random} initial conditions. For the first   initial conditions (first two rows),
an  oscillation is observed between two different states corresponding to the two branches
of Figure \ref{fig9}. This is confirmed by  the
corresponding probability distribution function~(PDF)
in which a well defined bimodal character can be identified(Fig.~\ref{fig11}~(a)).
The second random initial condition leads to a  different pattern~(see
 the last two rows of Fig.\ref{fig10}). The lattice is now divided into
two regions. In  each region, the system is   behaving  as before but there exists   a
relative phase lag between the two "sub-lattices". This feature is also visible in the PDF,
of Fig.\ref{fig11}~(b) where the new  bimodal PDF  is a
linear combination of the previous two distributions. As a
consequence, the value of $\overline{C(t)}$ is shifted towards "singular" points reported in
Fig.\ref{fig9}. These "frozen" states have extremely long and possibly infinite
relaxation times.

\begin{figure}[hbtp]
\begin{center}
\includegraphics[width=10cm]{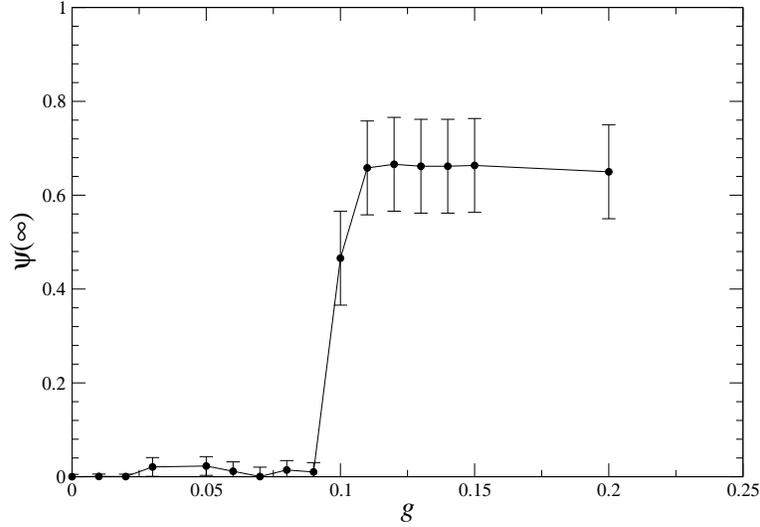}
\end{center}
\caption[]{Phase diagram for the undiluted logistic map: order
  parameter $\psi$ vs. coupling $g$ (see text). Here   size is
  $L=1024$, $\mu=1.7$, $T=10^4$.
\label{fig12}
}
\end{figure}

\noindent  Upon identifying a suitable order parameter and monitoring the
variation of such a parameter with coupling $g$, one finds a transition from a "high temperature" to a
"low temperature" phase, akin to that found for the Miller-Huse map.
A  proper definition of the order parameter is, however, a delicate matter.
Depending on the value of the coupling $g$, the spatial average $\overline{C(t)}$
is  a time constant or it follows a periodic sequence.
Hence, by computing
\begin{eqnarray}
\psi(t) &=& \biggl\vert\frac{\overline{C(t+1)}-\overline{C(t)}}{\overline{C_{ave}(t)}}\biggr\vert~~~~\hbox{with}~~~
\overline{C_{ave}(t)}=\frac{\left[\overline{C(t+1)}+\overline{C(t)}\right]}{2}.
\label{eq18}
\end{eqnarray}
for sufficiently long times, it is expected
to get value of $\psi(t \to \infty)$ close to $0$ for small $g$, and
a non-zero value for  $g\ge g_c$ \cite{Marq05}.  An abrupt change is indeed
observed (Fig. \ref{fig12}) for a given $g_c(\mu)$. For instance, one obtains  $g_c \sim 0.1$ for $\mu=1.7$.
More extensive and systematic calculations on this point are left for
a future study.

\subsection{Effect of a binary quenched disorder on the logistic
  coupled map lattice.}

\begin{figure}[hbtp]
\begin{center}
\includegraphics[width=10cm]{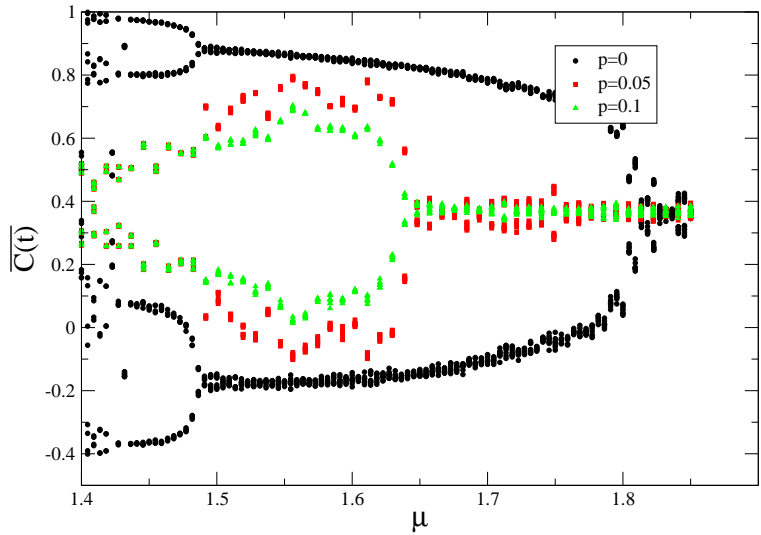}
\end{center}
\caption[]{(Color online) Phase diagram for the logistic CML with coupling $g=0.2$, size
$L=128$ with binary disorder  $p=0.05$ and $p=0.1$.
The ordered case $p=0$ is shown as a comparison. Spatial average
$\overline{C(t)} = \frac{1}{V} \sum_x C_x(t)$
is displayed as a function of  $\mu$ for  six  consecutive
times $T+1,T+2, T+3, T+4, T+5, T+6$ with $T=4.5 \times 10^5$.
\label{fig13}
}
\end{figure}

\noindent Let us now consider   the coupled logistic maps  in the presence of binary quenched disorder.
We find that even a small amount of disorder is sufficient
to modify  the ``conventional'' pattern reported on the regular
lattice, and this is true irrespective of   sizes $L$. Moreover, long transients are now present even in
the democratic coupling $g=0.2$. For instance we have explicitly
checked that, for $L=1024$, $T=10^4$ is no longer sufficient to reach the steady state.
We hence consider size $L=128$ for which a sufficiently long time ($T=4.5 \times 10^6$) can be
attained to ensure steady state conditions.
Spatial average $\overline{C(t)} $ is shown in figure~\ref{fig13}
versus parameter $\mu$,  for  coupling $g=0.2$ and   two   weak dilutions
$p=0.05$ and $p=0.1$. Once a small fraction of the total number of bonds is missing,
the formation of the period  doubling sequence is shifted to the left :  disorder
thus tends to inhibit the formation of the bifurcation structure.
An additional increase
in the dilution $p$   gives rise to a further shift  to the left, and eventually
to destruction of the collective dynamics  for a sufficiently high $p=p_c$.
This conclusion is supported by further calculations (not shown) on a smaller size
($L=64$) where a systematic trend to the left is observed as $p$ increases. However,
the large fluctuations present for such a small size
prevent  us from a quantitative measure of the critical threshold.

\subsection{The Lyapunov spectrum}

\noindent As suggested in~\cite{OHern96},  the
Lyapunov   spectrum is {\it a priori} another useful quantity
to analyze the collective
behavior of  CLM. The computation is performed by    recasting  Eq.(\ref{eq13}) in the
following form:
\begin{eqnarray}
C_{x}(t+1)&=& \sum_{y} M_{xy} F\left[C_{y}(t)\right]
\label{eq19}
\end{eqnarray}
where the matrix $M_{xy}$ stands for
\begin{eqnarray}
M_{xy} \left( \{W \}\right)&=& (1-\epsilon_x) \delta_{x y} + W_{xy} \delta_{|x-y|, 1}
\label{eq20}
\end{eqnarray}
Upon considering a variation $\delta C (t)$  of a given solution
$C^{0}(t) $, one obtains
\begin{eqnarray}
 \delta C_{x}(t+1)  &=&
  \sum_{y} M_{xy} F'\left[C^{0}_{y}(t)\right]
 \delta C_{y}(t)
\label{eq21}
\end{eqnarray}
The map    enters into  equation (\ref{eq21}) only through the value   $F'[C^{0}]$.
The Lyapunov exponents are   computed by taking into account the time evolution of an initially small
variation $\delta C_{x}(t=0)$. This procedure is standard and outlined
in~\cite{Benettin80}. To compute the
lower part of the spectrum with sufficient precision,  the
time window for the evolution is kept sufficiently small
and  the number of time intervals  for  normalization is increased  accordingly.
This procedure   has been applied  for both the Miller-Huse and the
logistic map. A cross-checking for the good performance of the
algorithm can be carried out by computing the exact eigenvalues for the uncoupled case $g = 0$
and the diffusion linear map . In this latter case, the eigenvalues of matrix
$M_{xy}$ are directly connected to the Lyapunov exponents.

\begin{figure}[hbtp]
\begin{center}
\includegraphics[width=10cm]{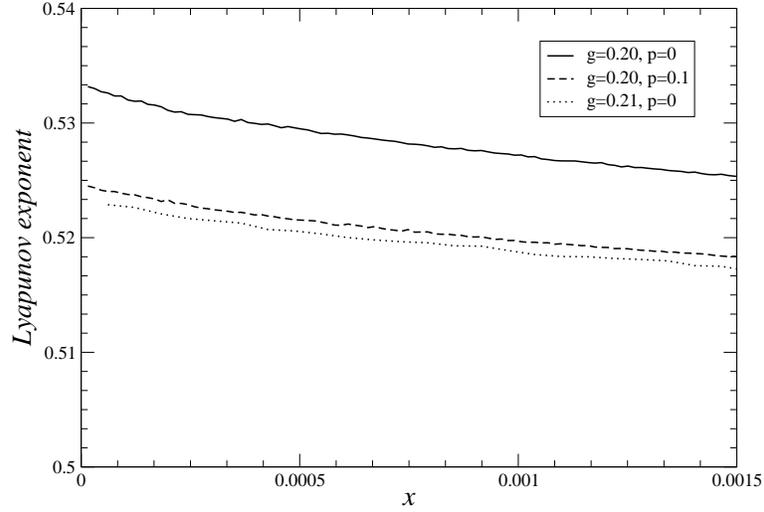}
\end{center}
\caption[]{The first $100$ Lyapunov exponents vs. the normalized index
$x=i/L^2$ for the two-dimensional  Miller-Huse CML with size $L=128$.
Full~(resp. dotted) line refers to a regular lattice at  $\mu=3$
and $g=0.20 < g_c$ (resp.$g=0.21> g_c$). The broken line refers to a diluted case
$g=0.20$ and $p=0.1$.
\label{fig14}
}
\end{figure}

\noindent In the regular case, the  Lyapunov exponent spectrum  has already been
carried out for  the Miller-Huse map at $\mu=3$ \cite{OHern96}.
In that work, it was suggested
that the Lyapunov spectrum is not sensitive to the presence of the ``ferromagnetic ordering''
discussed earlier. The independence of the spectrum
on the critical coupling $g_c$ was  also related to the fact that chaotic fluctuations,
 as measured by the Lyapunov spectrum, have  short range and thus decouple from the
the  long-range spatial effect giving rise to the ferromagnetic ordering at the
critical coupling $g_c$.
This result is confirmed by our computations which were carried  out
at various sizes (from $L=24$ as in~\cite{OHern96} to $L=256$). Figure~\ref{fig14} plots the
first 100 most unstable exponents (this represents a small fraction of the
spectrum whose size  is $V=L^2=16384$ for $L=128$)  versus the normalized index $x=i/L^2$ for
two values of the coupling (one below and the other above
the critical coupling $g_c \sim 0.205$).
As in \cite{OHern96}, no sign of transition can be traced in the results.
The spectrum is  continuous  starting with a value which is well
below the unique exponent of the uncoupled case $\sim 0.9$.
The independence of these results
for the upper spectrum on the size $L$ of the system has been verified.
The effect of the dilution on this
spectrum has been analyzed   in the presence
of different degrees $p$ of dilution. Fig. \ref{fig14} displays the
results for $p=0.1$ showing that disorder   lowers the values of the Lyapunov spectrum.

\begin{figure}[hbtp]
\begin{center}
\includegraphics[width=10cm]{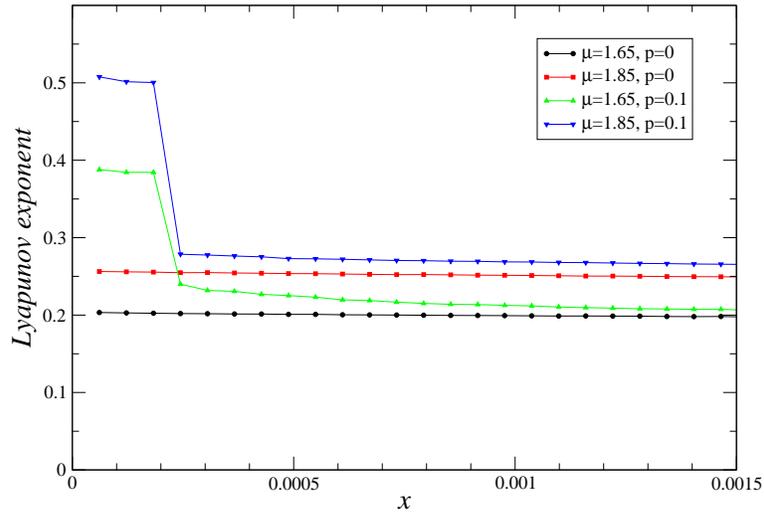}\\
\end{center}
\caption[]{(Color online) The first $100$ Lyapunov exponents vs. the normalized index  $x=i/L^2$
for the  logistic CML. Here the size is $L=128$,  coupling $g=0.2$
and two   parameters $\mu=1.65, 1.85$ are considered. Two curves refer to the
 the undiluted~($p=0$) and  diluted~($p=0.1$) logistic CML.
\label{fig15}
}
\end{figure}

\noindent For the logistic map, we are not aware of any previous calculations
 even in the regular  case. Fig. \ref{fig15} depicts our results both
in the undiluted and the diluted case in the case of a democratic coupling
$g=0.2$ and for two different values $\mu=1.85$ and $\mu=1.65$. For
 the first $\mu$
the spatial average $\overline{C(t)}$ is a time constant, whereas for the second value
of $\mu$, $\overline{C(t)}$ is time periodic. Nevertheless, no qualitavive difference is
found in the Lyapunov spectrum which is simply shifted to higher values as $\mu$ increases.
Unlike in the Miller-Huse case, the presence of disorder appears in the
 Lyapunov spectrum in the form of a gap separating the first few Lyapunov exponents
from the rest of the spectrum (Fig. \ref{fig15}).
Again, the effect of changing the value of the parameter $\mu$ is a shifting of the
numerical value of the exponents, in agreement with the undiluted
 case. On recalling the analogy between the Lyapunov spectrum and
 the problem of the Sch\"odinger equation on a lattice \cite{Isola90},
 the emergence of the gap in the lowest part of the Lyapunov spectrum
 in the presence of ``defects'', can be regarded as the analog of a
 ``localization'' mechanism. On the other hand this analogy is
 certainly limited, or it is at least incomplete, since it does not
 apply to  the Miller and Huse case
 discussed earlier.

\section{Conclusion}
\label{sect5}

\noindent We have shown that the phenomenology of spatio-temporal nonlinear  systems
may be significantly changed by a small amount of quenched disorder.
This feature has been demonstrated on  two-dimensional CML with
periodic boundary conditions. Two paradigmatic cases have been
investigated. The first one is the Miller and Huse case, which is known
to display a phase transition in the strong coupling (low temperature)
phase. The transition is found to be  delayed and even inhibited
by a small amount of binary noise. A percentage of
disconnected bonds as small as $10\%$ is sufficient to destroy the
ordered phase. This feature is mirrored by an analogous transition on
the persistence probability.
The second paradigmatic case concerns the logistic map  which
exhibits a remarkable time evolution of the spatial average
$\overline{C(t)}$ on a  regular
lattice. In this case,   issues regarding
the possibility of long transient behaviors  have been addressed. As it turns out, 
in the non-democratic case, the system might
get trapped into a metastable state away from the expected
bifurcation branch. In addition,   the phase
diagram $\overline{C(t)}$ versus $\mu$ is markedly affected (even in
the democratic case) by a small percentage of dilution~(e.g.~$5\%$).

\noindent This work is believed to unravel  some aspects of the interplay
between quenched disorder and non-linear processes for cases which were
not previously studied. In particular, it indicates that the lattice  structure affects in a major
way the system dynamics. Along this line of thought,
it would be worth modifying the nature of the disorder
  to understand the significance of  the qualitative nature of
disorder. This can be done, on  the one hand, by changing coupling amplitudes  with position rather
than by cutting some links as performed in the present paper, or, on
the other hand, by considering topological disorder such as a fractal stucture.

\begin{acknowledgments}
 Financial support from a CNR-CNRS exchange program is greatly acknowledged. We would like 
 to thank A. Torcini for useful criticisms and suggestions
as well as a careful reading of a first draft of the manuscript.
We have also benefited from useful suggestions from R. Livi, A. Politi, P. Marq and P. Manneville.
\end{acknowledgments}


\newpage

\begin{thebibliography}{910}
\bibitem{Morris93} S.W. Morris, E. Bodenschatz, D. S. Cannel, G. Ahlers, Phys. Rev. Lett. {\bf 71}, 2026 (1993)
\bibitem{Ouyang96} Q. Ouyang and J. M. Flesselles, Nature {\bf 379}, 143 (1996)
\bibitem{Lee96} K. J. Lee, E. C. Cox, R. E. Goldstein, Phys. Rev. Lett. {\bf 76}, 1174 (1996)
\bibitem{Qu97} Z. L. Qu, J. N. Weiss, A. Gartfinkel,
  Phys. Rev. Lett. {\bf 78}, 1387 (1997)
\bibitem{Kaneko93} K. Kaneko ed. {\it Theory and Applications of
  Coupled Map Lattice}  (Wiley, New York 1993)
\bibitem{Haus87}J. Haus and K. Kehr, Phys. Rep. {\bf 150}, 263 (1987); S. Havil and D. Ben-Avraham, Adv. Phys.{\bf 36}, 659 (1987); S. Alexander et al, Rev. Mod. Phys. {\bf 53}, 175 (1981);
J. P. Bouchaud and A. Georges, Phys. Rep. {\bf 195}, 127 (1990)
\bibitem{Giacometti96} A. Giacometti and KPN Murthy, Phys. Rev. E {\bf 53},
5647 (1996) and references therein.
\bibitem{Radons04} G. Radons, W. Just, P. Ha\"ussler (edts) {
\it Collective dynamics of non linear and disordered systems.}(Sringer
  Verlag, 2004)
\bibitem{Baroni01} L. Baroni, R. Livi and A. Torcini, Phys. Rev. E {\bf 63},
0362226 (2001); L. Angelini, M. Pellicoro and S. Stramaglia, Phys. Lett. A {\bf 285}, 293 (2001)
\bibitem{Miller93a} J. Miller and D.A. Huse, Phys. Rev. {\bf A}, 2528 (1993)
\bibitem{Lemaitre99} A. Lema\^itre and H. Chat\'e Phys. Rev. Lett. {\bf 82},
1140 (1999)
\bibitem{Lemaitre98} Ana\"el Lema\^itre, Ph.D thesis, \'Ecole
  Polytechnique, France (1998)
\bibitem{Stanley71} H.E. Stanley, "Introduction to Phase Transition and Critical Phenomena", Oxford University Press, New York 1971
\bibitem{Marq97} P. Marq, H. Chat\'e, and P. Manneville, Phys. Rev. E {\bf 55}, 2606
(1997)
\bibitem{Bray94}   Alan Bray Adv. Phys. 43, 357 (1994)
\bibitem{HohenbergHalperin} P. C. Hohenberg and B. I. Halperin Rev. Mod. Phys. 49, 435-479 (1977)
\bibitem{Krug} J. Krug and H. Spohn in "Solids Far from Equilibrium", C. Godr\`eche ed., Cambridge University Press, Cambridge 1991 pp. 479
\bibitem{Roder99} A. Roder, J. Adler and W. Janke, Physica A {\bf 265}, 28 (1999)
\bibitem{Marq05} A similar line of reasoning has already been followed by P. Marq (private communication)
\bibitem{OHern96} C. S. O'Hern, D. A. Egolf and H. S. Greenside,
Phys. Rev. E {\bf 53}, 3374 (1996)
\bibitem{Benettin80} G. Benettin, L. Galgani, A. Giorgilli, J-M. Strelcyn
Meccanica {\bf 15}, 9 (1980)
\bibitem{Isola90} S. Isola, A. Politi, S. Ruffo, A. Torcini,
  Phys. Lett. A {\bf 143}, 365 (1990)
\end{thebibliography}
\end{document}